\documentclass[iop]{emulateapj}

\newcommand{\kms}{\textrm{km~s$^{-1}$}}

\newcommand{\lsolar}{L$_{\odot}$}
\newcommand{\msolar}{M$_{\odot}$}
\newcommand{\ml}{M$_{\odot}$ yr$^{-1}$}
\newcommand{\mdot}{\dot{M}}

\usepackage{subfigure}
\usepackage{multirow}
\usepackage[dvips]{color}
\definecolor{Mygrey}{gray}{0.6}
 
\begin{document}

\title{An Excess of Mid-Infrared Emission from the Type Iax SN 2014dt}
\shorttitle{Spitzer Observations of SN 2014dt}
\author{Ori D. Fox\altaffilmark{1,2}, Joel Johansson \altaffilmark{3}, Mansi Kasliwal \altaffilmark{4}, Jennifer Andrews \altaffilmark{5}, John Bally \altaffilmark{6}, Howard E. Bond \altaffilmark{1,7}, Martha L. Boyer \altaffilmark{8,9}, R. D. Gehrz \altaffilmark{10}, George Helou \altaffilmark{11}, E. Y. Hsiao \altaffilmark{12}, Frank J. Masci \altaffilmark{11}, M. Parthasarathy \altaffilmark{13}, Nathan Smith \altaffilmark{5}, Samaporn Tinyanont \altaffilmark{4}, Schuyler D. Van Dyk \altaffilmark{11}}
\altaffiltext{1}{Space Telescope Science Institute, 3700 San Martin Drive, Baltimore, MD 21218, USA.}
\altaffiltext{2}{email: ofox@stsci.edu .}
\altaffiltext{3}{Benoziyo Center for Astrophysics, Weizmann Institute of Science, 76100 Rehovot, Israel}
\altaffiltext{4}{California Institute of Technology, Pasadena, CA 91125, USA.}
\altaffiltext{5}{Steward Observatory, 933 N. Cherry Ave., Tucson, AZ 85721, USA.}
\altaffiltext{6}{Center for Astrophysics and Space Astronomy, University of Colorado, Boulder, CO 80389, USA.}
\altaffiltext{7}{ Dept. of Astronomy \& Astrophysics, Pennsylvania State University, University Park, PA 16802 USA.}
\altaffiltext{8}{CRESST and Observational Cosmology Lab, Code 665, NASA Goddard Space Flight Center, Greenbelt, MD 20771 USA.}
\altaffiltext{9}{Department of Astronomy, University of Maryland, College Park, MD 20742 USA.}
\altaffiltext{10}{Minnesota Institute for Astrophysics, School of Physics and Astronomy, 116 Church Street S.E., University of Minnesota, Minneapolis, Minnesota 55455, U.S.A.}
\altaffiltext{11}{IPAC/Caltech, Mailcode 100-22, Pasadena, CA 91125, USA}
\altaffiltext{12}{Department of Physics, Florida State University, Tallahassee, FL 32306, USA}
\altaffiltext{13}{Indian Institute of Astrophysics, Bangalore 560034, India.}

\begin{abstract}

Supernovae Type Iax (SNe Iax) are less energetic and less luminous than typical thermonuclear explosions.   A suggested explanation for the observed characteristics of this subclass is a binary progenitor system consisting of a CO white dwarf primary accreting from a helium star companion.  A single-degenerate explosion channel might be expected to result in a dense circumstellar medium (CSM), although no evidence for such a CSM has yet been observed for this subclass.  Here we present recent {\it Spitzer}~observations of the SN Iax 2014dt obtained by the SPIRITS program nearly one year post-explosion that reveal a strong mid-IR excess over the expected fluxes of more normal SNe Ia.  This excess is consistent with $10^{-5}$~\msolar~of newly formed dust, which would be the first time that newly formed dust has been observed to form in a normal Type Ia.  The excess, however, is also consistent with a dusty CSM that was likely formed in pre-explosion mass-loss, thereby suggesting a single degenerate progenitor system.  Compared to other SNe Ia that show significant shock interaction (SNe Ia-CSM) and interacting core-collapse events (SNe IIn), this dust shell in SN 2014dt is less massive.  We consider the implications that such a pre-existing dust shell has for the progenitor system, including a binary system with a mass donor that is a red giant, a red supergiant, and an asymptotic giant branch star. 

\end{abstract}


\clearpage

\section{Introduction}
\label{sec:intro}

The ability to standardize Type Ia supernova (SN~Ia) light curves yields one of our most precise cosmological distance indicators \citep[e.g.,][]{phillips93}.  Despite these empirical relationships, questions remain about the underlying physics and progenitor systems.  The exploding primary is generally accepted to be a CO white dwarf (WD) that experiences a thermonuclear explosion, but the nature of the companion star remains ambiguous.  Evidence now exists for both single-degenerate and double-degenerate channels \citep[e.g.,][and references within]{maoz13}.

Recent studies reveal a subsample of SNe~Ia that are less energetic and less luminous than typical thermonuclear events \citep[e.g.,][and references within]{foley13,foley15}.  Designated the Type Iax subclass, the SN ejecta exhibit slower velocities than typical SNe Ia near maximum light.  In contrast to normal SNe Ia, which likely undergo a deflagration that transitions into a detonation \citep{khokhlov91}, the above characteristics of SNe Iax may suggest a full deflagration of a WD \citep{branch04,philips07}, although some SNe Iax are interpreted as Chandrasekhar-mass explosions experiencing pulsationally delayed detonations \citep{stritzinger15}.

There is growing evidence to suggest that slow Type Ia supernovae may be single degenerate. A UV pulse due to companion interaction was seen in iPTF14atg,  a SN2002es-like supernova \citep{cao15}.  Pre-explosion images of SN Iax 2012Z reveal a luminous blue source at the position of the SN, suggesting a non-degenerate He companion, although the data may also be interpreted as a massive primary star or an accretion disk around the exploding primary WD \citep{mccully14}.  Images of the SN Iax 2008ha four years post-explosion reveal a red source at the position of the SN that is consistent with either a thermally pulsating asymptotic giant branch (AGB) companion star or the bound remnant of the primary WD \citep{foley14}.  These results are used to argue that SNe Iax must have a diverse set of progenitors \citep[e.g.,][]{foley15, white15}.

\begin{figure*}
\begin{center}
\epsscale{1}
\plotone{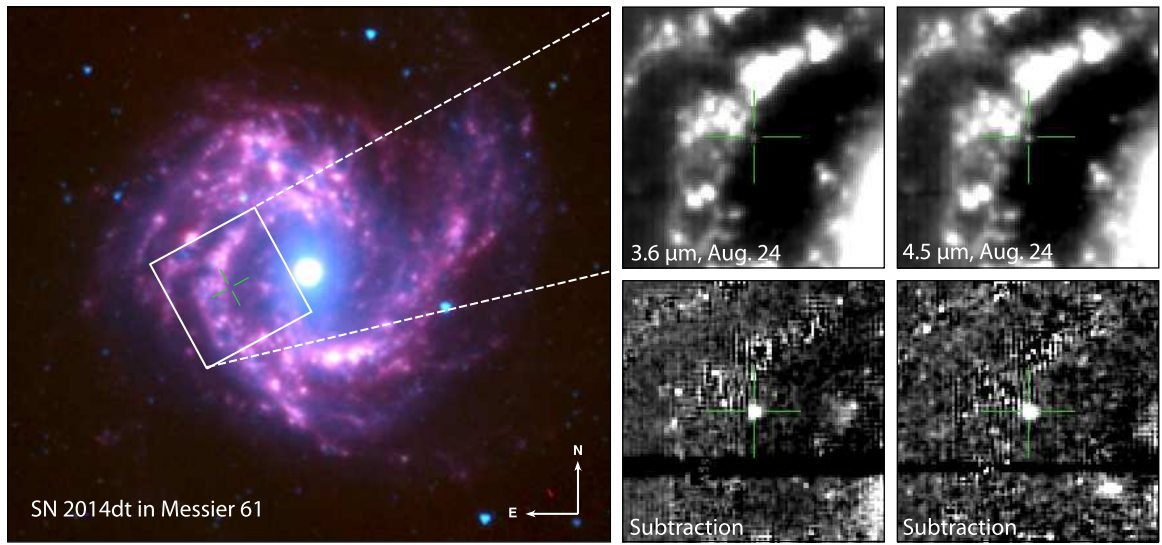}
\caption{Left panel: RGB composite of M\,61 using pre-SN Spitzer data (3.6, 4.5 and 8.0 $\mu$m) from 2004. The inset square and green cross indicates position of the right panel patches and SN 2014dt in M61. Right panels: $60\arcsec\times60\arcsec$ patches of the 2015 Aug. 24 SPIRITS 3.6 and 4.5 $\mu$m observations and subtractions of SN\,2014dt.
}
\label{f1}
\end{center}
\end{figure*}

The SN Iax 2014dt was discovered on 2014 Oct. 29.838 in M61 \citep{nakano14,ochner14}.  Pre-explosion imaging places upper-limits on the progenitor that are consistent with a CO white-dwarf primary and a non-degenerate He companion progenitor system, as was suggested for SN 2012Z, although perhaps with a slightly smaller and/or hotter donor \citep{foley15}.  The pre-explosion data, however, are also consistent with a low-mass red giant (RG) or main-sequence star companion.  If detected, a circumstellar medium (CSM) may be able to offer clues about the SN 2014dt progenitor system and pre-explosion mass-loss.  For example, the presence of H-rich and dense CSM seen in the Type Ia-CSM subclass suggests a non-degenerate companion star in the progenitor system \citep[e.g.,][]{silverman13b,fox15}.  Narrow lines, often taken to be a key signature of a dense CSM, have not been previously observed in SNe Iax, but this result may be due to a lower density CSM or slower shock velocity.  Like SNe Ia-CSM and their core-collapse counterparts (SNe IIn), another approach to detecting the presence of a CSM is through the presence of a mid-infrared (mid-IR) excess of thermal emission resulting from warm dust, in some cases years post-explosion \citep[e.g.,][]{fox11, fox13b}.  

In this {\it Letter} we present $Spitzer~Space~Telescope$~\citep{Werner:2004p24304, Gehrz:2007p24308} data of SN 2014dt obtained more than 300 days post-maximum by the SPitzer InfraRed Intensive Transients Survey (SPIRITS; \citealt{Kasliwal:2014p24264}).  Section \ref{sec:2} lists the details of the observations; $Spitzer$ photometry constrains the dust mass and temperature, and thus the luminosity. We explore the origin and heating mechanism of the dust in \S \ref{sec:3}. Section \ref{sec:4} presents our conclusions.

\section{Observations}
\label{sec:2}

\subsection{Warm $Spitzer$/IRAC Photometry}
\label{sec:irac}

Table \ref{tab1} summarizes observations of SN 2014dt made by SPIRITS.  This survey provides a systematic transient search of 194 nearby galaxies within 20\,Mpc, on timescales ranging between a day to a year, to a depth of 20 mag in the two {\it Spitzer}/IRAC \citep{fazio04} channels at 3.6 and 4.5 \micron.  SPIRITS is an exploration science program that has been awarded 1130\,hours of {\it Spitzer} time over three years (2014--2016; PID \#11063 PI Kasliwal). Concomitantly, the SPIRITS team undertakes extensive ground-based monitoring of these galaxies in the optical and near-infrared.  

The {\it Spitzer} Heritage Archive (SHA)\footnote{SHA can be accessed from http://sha.ipac.caltech.edu/applications/Spitzer/SHA/ .} provides access to the Post Basic Calibrated Data ({\tt pbcd}), which are already fully coadded and calibrated.  Figure \ref{f1} shows a RGB image of combined 3.6, 4.5 and 8.0 \micron~images at a single pre-SN epoch.  The SN host galaxies tend to be bright and exhibit background-flux variations on small spatial scales.  SPIRITS implements template subtraction to reduce photometric confusion from the underlying galaxy \footnote{http://web.ipac.caltech.edu/staff/fmasci/home/mystats/ApPhotUncert.pdf}.  Forced aperture photometry is then performed on the stack of IRAC co-add images for each channel. An aperture of fixed radius 4 pixels~is centered on the position of SN2014dt.  An aperture correction of 1.21$\times$~is applied to the total flux in the aperture for both channels. The background is computed using a median in an annulus with inner/outer radii of 8\arcsec/15\arcsec.  Flux errors account for Poisson noise from the source and uncertainties in the local background estimate.  Table \ref{tab1} lists and Figure \ref{f3} plots the resulting photometry.  Throughout the paper we use the zero magnitude fluxes for the IRAC Channels 1 and 2 (CH1 and CH2, with central wavelengths of 3.6 and 4.5 $\mu$m, respectively) of $F_{\nu,0}^{\rm CH1} = 280.9$\,Jy and $F_{\nu,0}^{\rm CH2} = 179.7$\,Jy.

\begin{deluxetable*}{ l c c c c c c}
\tablewidth{0pt}
\tabletypesize{\footnotesize}
\tablecaption{$Spitzer$~Observations and IR~Fitting Parameters ($a = 0.1$ \micron~amorphous carbon)\label{tab1}}
\tablecolumns{7}
\tablehead{
\colhead{JD} & \colhead{Epoch} & \colhead{3.6~\micron\tablenotemark{1}} & \colhead{4.5~\micron\tablenotemark{1}} & \colhead{$M_{\rm d}$ (\msolar)} & \colhead{$T_{\rm d}$} (K) & \colhead{$L_{\rm d}$ (\lsolar)}\\
\colhead{$-$2,450,000}&\colhead{(days)}& \multicolumn{2}{c}{($\mu$Jy)} & \colhead{ } & \colhead{ } &\colhead{ } 
}
\startdata
7259 & 309 & 49 (16) & 52 (12) & $1.35\times10^{-5}$ &  711   &  4.75$\times10^{5}$\\
7267 & 317 & 52 (18) & 58 (13) &$1.83\times10^{-5}$  &  679     &  4.98$\times10^{5}$\\
7286 & 336 & 75 (18) & 73 (14) &$1.33\times10^{-5}$  &  770     &  7.48$\times10^{5}$
\enddata
\tablenotetext{1}{1$\sigma$~uncertainties are given in parentheses.}
\end{deluxetable*}

\subsection{Optical and Near-IR Photometry}
\label{sec:uvir}

Optical photometry were obtained with the Las Cumbres Observatory Global Telescope (LCOGT) Network (PIs: J. Bally, E. Gomez and K. Finkelstein) in $BVR$ and $ri$ filters. The data were reduced and analyzed with standard IRAF\footnote{IRAF: The Image Reduction and Analysis Facility is distributed by the National Optical Astronomy Observatory, which is operated by the Association of Universities for Research in Astronomy (AURA) under cooperative agreement with the National Science Foundation (NSF).} routines, using the QUBA pipeline \citep[see][for details]{2011MNRAS.416.3138V}.  Template galaxy subtraction is {\it not} performed.  The SN magnitudes are measured with a point-spread function (PSF) fitting technique (using daophot) and calibrated using SDSS photometry of stars in the field. For the light curve analysis we also include $gri$ and $R$-band photometry from the Palomar 48 and 60-inch telescopes, $UBV$ photometry\footnote{made public by Peter Brown on the {\it Swift}~Supernovae page} from the Ultra-Violet/Optical Telescope \citep[UVOT,][]{2005SSRv..120...95R} on the Swift spacecraft \citep{2004ApJ...611.1005G} and measurements by amateurs \footnote{made public on the AAVSO and the Bright Supernovae pages}. In order to match the SN 2014dt photometry, the $RI$ data were converted to $ri$ magnitudes using the conversions in \citet{2006A&A...460..339J}.

Near-IR (NIR) observations in the Mauna Kea Observatory $JHK_s$ filters were carried out with the United Kingdom Infrared Telescope (UKIRT). Template galaxy subtraction is {\it not} performed.  PSF fitting photometry was performed on the sky-subtracted frames, calibrated using 2MASS stars in the field. We also include the early NIR photometry from \citet{joshi14}.

\subsection{Distance to M61}
\label{sec:distance}

\begin{figure}
\begin{center}
\epsscale{1.2}
\hspace{-0.1in}
\plotone{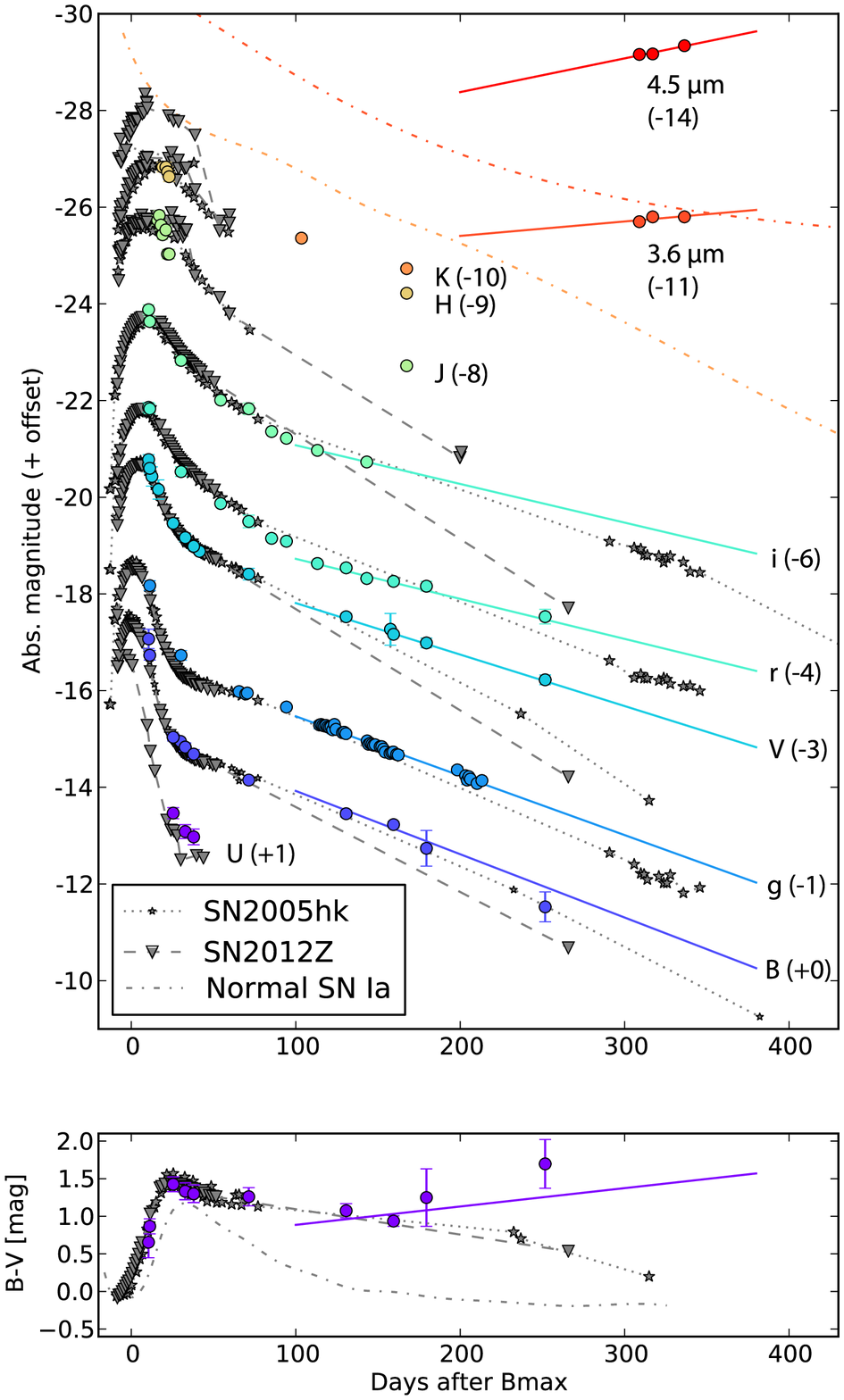}
\caption{Upper panel: Optical, near- and mid-IR light curves of SN\,2014dt (colored circles and solid lines) assuming a distance modulus $\mu=31.43$ mag and $MJD_{B, {\rm max}}=56950$. 
The light curves of SNe 2005hk (grey stars and dotted lines; \citealt{philips07,2008ApJ...680..580S}) and 2012Z (grey triangles and dashed lines; \citealt{stritzinger15,2015ApJ...806..191Y}) are shown for comparison as well as the 3.6 and 4.5  $\mu$m templates for normal SNe Ia (dash-dotted lines) from Johansson et al. (2014). Bottom panel: B-V color evolution of SN\,2014dt compared to a normal SN Ia and SNe Iax 2005hk and 2012Z.
}
\label{f3}
\end{center}
\end{figure}

Figure \ref{f3} plots the multi-wavelength light curve of SN 2014dt compared to other SNe Iax 2005k and 2012Z.  We note that SN 2014dt was discovered after peak brightness, but the light curves indicate that it peaked around Oct. 20 (MJD=56950).  This explosion date is comparable to the time of maximum deduced from spectral cross correlations \citep{ochner14,foley15}.  Two different methods based on the Type II SN 2008in \citep{2011ApJ...736...76R} yield distances to M61 ranging between a distance modulus $\mu = 30.45$ mag (12.3 Mpc, EPM method, \cite{2014ApJ...782...98B}) and $\mu = 31.43$ mag (19.3 Mpc, Photospheric magnitude method, \cite{2014AJ....148..107R}).  While the former is consistent with the Tully-Fisher estimate \citep{1997A&A...323...14S} and used to put limits on the progenitor system in \citet{foley15}, the larger distance makes SN 2014dt appear to have similar absolute magnitudes as SNe 2005hk and 2012Z, peaking at $M_V \sim -18$ mag.  We assume a distance of 19.3 Mpc for the analysis in this paper.

\begin{figure*}
\begin{center}
\plotone{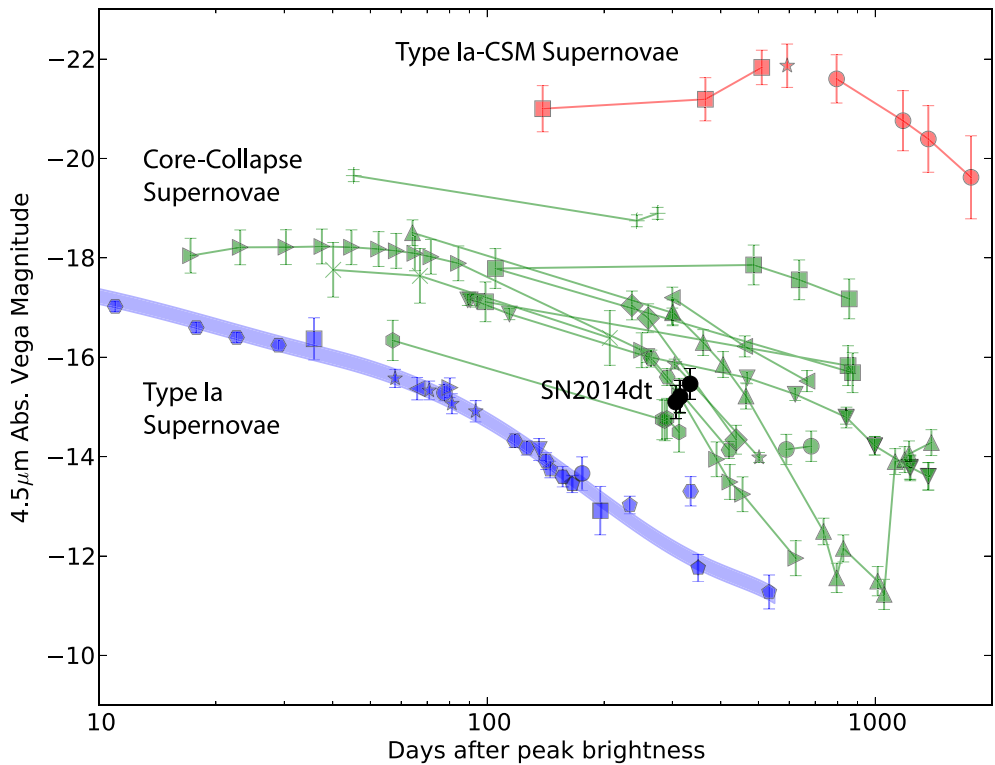}
\caption{Collage of 4.5 $\mu$m absolute magnitudes of normal SNe Ia (blue symbols, from \citealt{johansson14}), core-collapse SNe (green symbols, from Tinyanont et al. {\it in prep.}, \citealt{szalai13}, and references therein) and strongly interacting SNe Ia-CSM (red symbols, from \citealt{fox11} and \citealt{fox13b}).}
\label{f4}
\end{center}
\end{figure*}

\subsection{A Mid-IR Excess}
\label{sec:excess}

Both SN 2012Z  \citep{stritzinger15,2015ApJ...806..191Y} and SN 2005hk \citep{philips07,2008ApJ...680..580S} had low host-galaxy extinction $E(B-V) \leq 0.10$, and SN 2014dt seems to follow the early color evolution of both these SNe. However, at around 100 days post peak SN 2014dt starts to exhibit redder colors than SNe 2005hk and 2012Z. As we explain below, the redder colors likely stem from additional flux at longer wavelengths, rather than dust attenuation of bluer light.

Figure \ref{f3} shows that SN 2014dt has plateaued between days 298 and 326.  Such behavior is not typical of thermonuclear SNe Ia, which follow radioactive decay rates at these epochs, even in the mid-IR \citep{johansson14}.  For a comparison, Figure \ref{f4} plots the SN 2014dt {\it Spitzer}~Channel 2 photometry over a mid-IR template compiled for both normal SNe Ia \citep{johansson14} and a sample of SNe with observed dust present.  SN 2014dt shows a clear mid-IR excess in both {\it Spitzer}~bands.  

At late times, SN Ia light curves transition from a regime dominated by gamma-rays to one dominated by positrons.  In the case of energy deposition by positrons, the IR light curve slope can change due to various physical effects that govern the positron escape probability \citep[e.g., magnetic fields;][]{Penney:2014p24317}. For a normal SN Ia at 300 days, this effect is only on the order of a tenth of a magnitude. We note that even if the positrons are fully trapped, this effect cannot explain the observed IR plateau.

The source of the mid-IR excess is therefore likely to be warm dust.  {\it Spitzer}~photometry offers the advantage of spanning the peak of the blackbody produced by warm grains.  The flux can therefore be fit as a function of the dust temperature, $T_{\rm d}$, and mass, $M_{\rm d}$,
\begin{equation}
\label{eqn:flux2}
F_\nu = \frac{M_{\rm d} B_\nu(T_{\rm d}) \kappa_\nu(a)}{d^2},
\end{equation}
assuming optically thin dust with particle radius $a$, at a distance $d$ from the observer, and thermally emitting at a single equilibrium temperature \citep[e.g.,][]{hildebrand83}, where $\kappa_\nu(a)$ is the dust absorption coefficient.  

\begin{figure}
\begin{center}
\epsscale{1.2}
\plotone{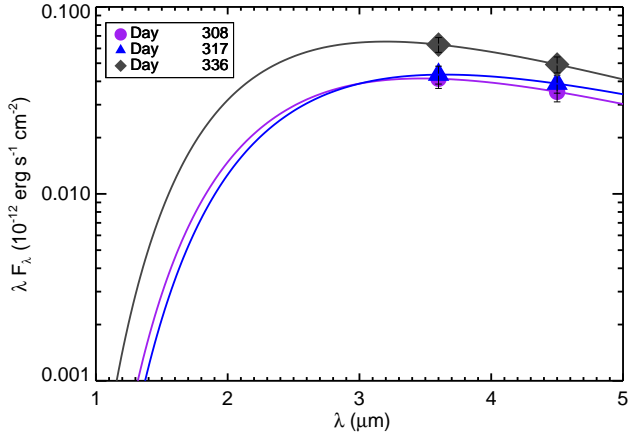}
\caption{Photometry of SNe 2014dt in $Spitzer$/IRAC Channels 1 (3.6~\micron) \&~2 (4.5 \micron).  Overplotted are the resulting best fits of Equation \ref{eqn:flux2}.
}
\label{f2}
\end{center}
\end{figure}

The dust composition and temperature distribution is unknown.  Given only 2 photometry points, however, we assume a simple dust population of a single size and temperature composed entirely of amorphous carbon (AC; \citealt{williams15}).  These assumptions are consistent with previous analysis of warm {\it Spitzer} data \citep[e.g.,][]{fox11,fox13b}, which allows for meaningful comparisons.  Figure \ref{f2} shows the best fit of Equation \ref{eqn:flux2} obtained with the IDL {\tt MPFIT} function \citep{markwardt09}, which minimizes the value of $\chi^2$ by varying $M_{\rm d}$ and $T_{\rm d}$.  With only two photometry points at each epoch, we limit our fits to a single component (see Figure \ref{f2}).  Table \ref{tab1} lists the best-fit parameters for AC grains of size $a = 0.1$ \micron.   The dust masses presented here are likely lower limits since a bulk of the dust likely resides at cooler temperatures not probed by {\it Spitzer}.  For a distance to M61 of only $\sim$12 Mpc, however, both the dust mass and luminosity presented in Table \ref{tab1} would decrease by a factor of $\sim2.5\times$, but the mid-IR excess relative to the normal SN Ia sample would still be present.

Compared to the mid-IR excess observed in the SNe Ia-CSM 2002ic and 2005gj \citep[e.g.][]{fox13b}, the dust luminosity (and mass) of SN 2014dt is approximately two orders of magnitude smaller than the SNe Ia-CSM, while the temperatures are about the same (but note the fit is made with only two photometry points).  We explore the implications of these similarities and differences below.

\section{Analysis and Discussion}
\label{sec:3}

\subsection{Possible Origins and Heating Mechanisms}
\label{sec:origin}

The source of the mid-IR emission is likely warm dust, but the origin and heating mechanism of the dust are less clear.  The dust may be either newly formed or pre-existing, and either shock or radiatively heated; see \citet{fox10} for a full discussion.  To discriminate between possible scenarios, we first assume a spherically symmetric, optically thick dust shell and calculate the blackbody radius, $r_{\rm bb} = [L_d/(4 \pi \sigma T_d^4)]^{1/2}$,
which sets a {\it minimum} shell size.  The measured luminosity $L_d \approx 5\times10^{5}$~\lsolar\ and dust temperature $T_{\rm d} \approx 700$~K yield a blackbody radius of $r_{\rm bb} \approx 3.5\times10^{15}$~cm.  A caveat that should be considered is that the dust is likely optically thin since SN 2014t does not exhibit much extinction, so the dust shell radius is likely larger than the calculated blackbody radius or the dust is asymmetrically distributed.

In any case, the blackbody radius is comparable to the distance traveled by material with a time averaged velocity of $\sim1300$~\kms~over $\sim$300 days.  Indeed, models suggest dust can form in SNe Ia at these velocities, timescales, and total mass \citep{nozawa11}.  Furthermore, SNe Iax are defined by their low velocities \citep[e.g.,][and references within]{foley13}.  It is interesting to note that no other SN Ia has ever been observed to form new dust (and this paper presents the first late-time mid-IR observations of any SN Iax).  Simultaneous optical and IR data would be beneficial since newly formed dust is often associated with not only a rise in the IR luminosity, but also a simultaneous drop in the visual luminosity due to absorption.  Unfortunately, our optical photometry do not extend to the epochs of the {\it Spitzer}~observations.

Another possibility that is consistent with the data is pre-existing dust in the CSM.  Figure 8 in \citet{fox10} shows that the minimum radius derived above is approximately consistent with the distance to which such dust would have been vaporized (assuming a vaporization temperature of $\sim$2000 K) by the peak luminosity of SN 2014dt ($M\approx-18$~mag; \citealt{foley14}).  A dust shell lying at the vaporization radius is usually a signature of continuous mass loss from the progenitor primary (or companion) since a temporary mass-loss scenario (i.e., eruption) would have to include a contrived timeline.  For the case of pre-existing dust, the heating mechanism could be an IR echo, shocks, or radiation from the X-ray/UV/optical emission produced by shock interaction at an inner radius.  An IR echo at this radius would last only a few days ($t \approx \frac{2r}{c}$), so the duration of the mid-IR light curve rules out that scenario.  Shock heating of the dust, however, is possible given that the $\sim$1300~\kms~shock radius corresponds to both the derived blackbody and vaporization radii.

Alternatively, radiative heating of a pre-existing dust shell by X-ray/UV/optical emission generated by the shock interaction has been proposed to explain the late-time IR emission observed in SNe~IIn and Ia-CSM \citep[e.g.,][]{fox11,fox13,fox13b}.  Assuming an optically thin dust shell, the observed dust temperature ($T_{\rm d}$) and shell radius ($r_{\rm d}$) require a combined optical, ultraviolet, and/or X-ray flux given by
\begin{equation}
\label{eqn:lbol}
L_{\rm opt/UV/X}  = \frac{64}{3} \rho a r_{\rm d}^2 \sigma T_{\rm SN}^4 \frac{\int{B_\nu (T_{\rm d}) \kappa(\nu) d\nu}}{\int{B_\nu(T_{\rm SN}) Q_{\rm abs}(\nu) d\nu}} 
\end{equation}
for a dust bulk (volume) density $\rho$ and an effective SN blackbody temperature $T_{\rm SN}$, where $Q_{\rm abs}$ is the dust absorption efficiency, and $\kappa(\nu)$ is the dust absorption coefficient.  Figure 4 in \citet{fox13} shows that the blackbody radius of SNe 2014dt [$r_{\rm bb} \approx$ 3$\times 10^{15}$~cm] at temperatures $T_{\rm d} \approx 700$~K require optical and/or X-ray luminosities in the range $L_{\rm opt/UV/X} \approx 10^7$~\lsolar.  Optical/UV/and X-ray observations could confirm the predicted luminosities.  Furthermore, narrow lines in the optical spectra would offer clear evidence of CSM interaction.  As already noted, our optical photometry do not extend to the epochs of the {\it Spitzer}~observations, and we have no optical spectroscopy at these later epochs to search for narrow lines (the original classification spectrum does not show evidence of narrow lines at early times; \citealt{foley15}).

\subsection{Progenitor Mass-Loss Rate}
\label{sec:massloss}

Mid-IR wavelengths probe the characteristics of the CSM at the dust-shell radius.  Assuming a dust-to-gas mass ratio expected in the H-rich envelope of a massive star, $Z_{\rm d} = M_{\rm d}/M_{\rm g} \approx 0.01$, the dust-shell mass can be tied to the progenitor's total mass-loss rate, 
\begin{eqnarray}
\label{eqn:ml}
\mdot & = & \frac{M_{\rm d}}{Z_{\rm d} \Delta r} v_{\rm w} \nonumber \\ 
& = & \frac{3}{4} \Big(\frac{M_{\rm d}}{\rm M_{\odot}}\Big) \Big(\frac{v_{\rm w}}{120~\rm km~s^{-1}}\Big) \Big(\frac{5 \times 10^{16}~\rm cm}{r}\Big) \Big(\frac{r}{\Delta r}\Big) [{\rm M}_{\odot}~{\rm yr}^{-1}],
\end{eqnarray}
for a progenitor wind speed $v_{\rm w}$.  Since we do not know the pre-SN wind speed, we assume a speed of $\sim$10 \kms.  If we also assume a thin shell, ${\Delta r}/r = 1/10$, Equation \ref{eqn:ml} yields a mass-loss rate of $<10^{-6}~(v_{\rm w}/10~{\rm \kms})$~\ml, which is only an upper limit since the blackbody radius is only a lower limit.  This mass-loss rate is consistent with either a RG or red supergiant (RSG) \citep{Gehrz:1971aa,drout15} or an AGB star \citep{marshall04}.

\section{Summary}
\label{sec:4}

This {\it Letter} presents new $Spitzer$~data on the SN Iax 2014dt obtained nearly one year post-discovery.  The mid-IR data show evidence of emission from warm dust.  The warm-dust parameters are consistent with both newly formed dust and a pre-existing dust shell that is either heated by the forward shock or radiatively heated by optical, UV, and X-ray emission generated by the shock at interior radii.  In the former case, newly formed dust may be suggestive of a core-collapse origin \citep{valenti09}.  In the latter case, the pre-existing CSM suggests a non-degenerate companion star.  This CSM has a derived mass-loss rate of $<10^{-6}$~\ml.  Such a mass-loss rate is nearly a factor 10$\times$~smaller than the Type Ia-CSM PTF11kx \citep{dilday12}, which may explain both the lack of absorption and shock interaction observed in the earliest spectrum \citep{foley15}.  Without knowledge of the pre-SN wind speed, these mass-loss rates have a number of degeneracies when it comes to the progenitor system, including a RG, RSG, and even a AGB star.  Although not considered in the original paper \citep{foley14}, such a dusty CSM may also be the source of the red emission observed in the vicinity of SN 2008ha.  While pre-SN mass loss often suggests a single-degenerate channel, a caveat should be noted that such mass-loss rates have also been derived for both the core-degenerate and double-degenerate models \citep[e.g.,][]{soker13,shen13}.  Future multi-wavelength observations of SNe Iax, and SN 2014dt in particular, will be necessary to disentangle the various scenarios.\\

\vspace{10 mm}

This work is based on data obtained via Program \#11063 with the {\it Spitzer Space Telescope}, which is operated by the Jet Propulsion Laboratory, California Institute of Technology, under a contract with NASA. Support for this work was provided by NASA through an award issued by JPL/Caltech.  The authors thank Ryan Foley for useful discussions.  We thank Peter Milne for UKIRT observations.

\bibliographystyle{apj2}
\bibliography{references}

\begin{thebibliography}{}
\expandafter\ifx\csname natexlab\endcsname\relax\def\natexlab#1{#1}\fi

\bibitem[{{Bose} \& {Kumar}(2014)}]{2014ApJ...782...98B}
{Bose}, S., \& {Kumar}, B. 2014, \apj, 782, 98

\bibitem[{Branch {et~al.}(2004)Branch, Baron, Thomas, Kasen, Li, \&
  Filippenko}]{branch04}
Branch, D., Baron, E., Thomas, R.~C., {et~al.} 2004, PASP, 116, 903

\bibitem[{Cao {et~al.}(2015)Cao, Kulkarni, Howell, Gal-Yam, Kasliwal, Valenti,
  Johansson, Amanullah, Goobar, Sollerman, Taddia, Horesh, Sagiv, Cenko,
  Nugent, Arcavi, Surace, Wo{\'z}niak, Moody, Rebbapragada, Bue, \&
  Gehrels}]{cao15}
Cao, Y., Kulkarni, S.~R., Howell, D.~A., {et~al.} 2015, Nature, 521, 328

\bibitem[{Dilday {et~al.}(2012)Dilday, Howell, Cenko, Silverman, Nugent,
  Sullivan, Ben-ami, Bildsten, Bolte, Endl, Filippenko, Gnat, Horesh, Hsiao,
  Kasliwal, Kirkman, Maguire, Marcy, Moore, Pan, Parrent, Podsiadlowski,
  Quimby, Sternberg, Suzuki, Tytler, Xu, Bloom, Gal-Yam, Hook, Kulkarni, Law,
  Ofek, Polishook, \& Poznanski}]{dilday12}
Dilday, B., Howell, D.~A., Cenko, S.~B., {et~al.} 2012, Science, 337, 942

\bibitem[{Drout {et~al.}(2015)Drout, Milisavljevic, Parrent, Margutti, Kamble,
  Soderberg, Challis, Chornock, Fong, Frank, Gehrels, Graham, Hsiao, Itagaki,
  Kasliwal, Kirshner, Macomb, Marion, Norris, \& Phillips}]{drout15}
Drout, M.~R., Milisavljevic, D., Parrent, J., {et~al.} 2015, arXiv: 1507.02694

\bibitem[{Fazio {et~al.}(2004)Fazio, Hora, Allen, Ashby, Barmby, Deutsch,
  Huang, Kleiner, Marengo, Megeath, Melnick, Pahre, Patten, Polizotti, Smith,
  Taylor, Wang, Willner, Hoffmann, Pipher, Forrest, McMurty, McCreight,
  McKelvey, McMurray, Koch, Moseley, Arendt, Mentzell, Marx, Losch, Mayman,
  Eichhorn, Krebs, Jhabvala, Gezari, Fixsen, Flores, Shakoorzadeh, Jungo,
  Hakun, Workman, Karpati, Kichak, Whitley, Mann, Tollestrup, Eisenhardt,
  Stern, Gorjian, Bhattacharya, Carey, Nelson, Glaccum, Lacy, Lowrance, Laine,
  Reach, Stauffer, Surace, Wilson, Wright, Hoffman, Domingo, \&
  Cohen}]{fazio04}
Fazio, G.~G., Hora, J.~L., Allen, L.~E., {et~al.} 2004, ApJS, 154, 10

\bibitem[{Foley {et~al.}(2014)Foley, McCully, Jha, Bildsten, Fong, Narayan,
  Rest, \& Stritzinger}]{foley14}
Foley, R.~J., McCully, C., Jha, S.~W., {et~al.} 2014, ApJ, 792, 29

\bibitem[{Foley {et~al.}(2015)Foley, van Dyk, Jha, Clubb, Filippenko, Mauerhan,
  Miller, \& Smith}]{foley15}
Foley, R.~J., van Dyk, S.~D., Jha, S.~W., {et~al.} 2015, ApJL, 798, L37

\bibitem[{Foley {et~al.}(2013)Foley, Challis, Chornock, Ganeshalingam, Li,
  Marion, Morrell, Pignata, Stritzinger, Silverman, Wang, Anderson, Filippenko,
  Freedman, Hamuy, Jha, Kirshner, McCully, Persson, Phillips, Reichart, \&
  Soderberg}]{foley13}
Foley, R.~J., Challis, P.~J., Chornock, R., {et~al.} 2013, ApJ, 767, 57

\bibitem[{Fox {et~al.}(2010)Fox, Chevalier, Dwek, Skrutskie, Sugerman, \&
  Leisenring}]{fox10}
Fox, O.~D., Chevalier, R.~A., Dwek, E., {et~al.} 2010, ApJ, 725, 1768

\bibitem[{Fox \& Filippenko(2013)}]{fox13b}
Fox, O.~D., \& Filippenko, A.~V. 2013, ApJL, 772, L6

\bibitem[{Fox {et~al.}(2013)Fox, Filippenko, Skrutskie, Silverman,
  Ganeshalingam, Cenko, \& Clubb}]{fox13}
Fox, O.~D., Filippenko, A.~V., Skrutskie, M.~F., {et~al.} 2013, AJ, 146, 2

\bibitem[{Fox {et~al.}(2011)Fox, Chevalier, Skrutskie, Soderberg, Filippenko,
  Ganeshalingam, Silverman, Smith, \& Steele}]{fox11}
Fox, O.~D., Chevalier, R.~A., Skrutskie, M.~F., {et~al.} 2011, ApJ, 741, 7

\bibitem[{Fox {et~al.}(2015)Fox, Silverman, Filippenko, Mauerhan, Becker,
  Borish, Cenko, Clubb, Graham, Hsiao, Kelly, Lee, Marion, Milisavljevic,
  Parrent, Shivvers, Skrutskie, Smith, Wilson, \& Zheng}]{fox15}
Fox, O.~D., Silverman, J.~M., Filippenko, A.~V., {et~al.} 2015, MNRAS, 447, 772

\bibitem[{{Gehrels} {et~al.}(2004){Gehrels}, {Chincarini}, {Giommi}, {Mason},
  {Nousek}, {Wells}, {White}, {Barthelmy}, {Burrows}, {Cominsky}, {Hurley},
  {Marshall}, {M{\'e}sz{\'a}ros}, {Roming}, {Angelini}, {Barbier}, {Belloni},
  {Campana}, {Caraveo}, {Chester}, {Citterio}, {Cline}, {Cropper}, {Cummings},
  {Dean}, {Feigelson}, {Fenimore}, {Frail}, {Fruchter}, {Garmire}, {Gendreau},
  {Ghisellini}, {Greiner}, {Hill}, {Hunsberger}, {Krimm}, {Kulkarni}, {Kumar},
  {Lebrun}, {Lloyd-Ronning}, {Markwardt}, {Mattson}, {Mushotzky}, {Norris},
  {Osborne}, {Paczynski}, {Palmer}, {Park}, {Parsons}, {Paul}, {Rees},
  {Reynolds}, {Rhoads}, {Sasseen}, {Schaefer}, {Short}, {Smale}, {Smith},
  {Stella}, {Tagliaferri}, {Takahashi}, {Tashiro}, {Townsley}, {Tueller},
  {Turner}, {Vietri}, {Voges}, {Ward}, {Willingale}, {Zerbi}, \&
  {Zhang}}]{2004ApJ...611.1005G}
{Gehrels}, N., {Chincarini}, G., {Giommi}, P., {et~al.} 2004, \apj, 611, 1005

\bibitem[{Gehrz \& Woolf(1971)}]{Gehrz:1971aa}
Gehrz, R.~D., \& Woolf, N.~J. 1971, ApJ, 165, 285

\bibitem[{Gehrz {et~al.}(2007)Gehrz, Roellig, Werner, Fazio, Houck, Low, Rieke,
  Soifer, Levine, \& Romana}]{Gehrz:2007p24308}
Gehrz, R.~D., Roellig, T.~L., Werner, M.~W., {et~al.} 2007, Review of
  Scientific Instruments, 78, 1302

\bibitem[{Hildebrand(1983)}]{hildebrand83}
Hildebrand, R.~H. 1983, QJRAS, 24, 267

\bibitem[{Johansson {et~al.}(2014)Johansson, Goobar, Kasliwal, Helou, Masci,
  Tinyanont, Jencson, Cao, Fox, Kromer, Amanullah, Banerjee, Joshi, Jerkstrand,
  Kankare, \& Prince}]{johansson14}
Johansson, J., Goobar, A., Kasliwal, M.~M., {et~al.} 2014, arXiv:1411.3332

\bibitem[{{Jordi} {et~al.}(2006){Jordi}, {Grebel}, \&
  {Ammon}}]{2006A&A...460..339J}
{Jordi}, K., {Grebel}, E.~K., \& {Ammon}, K. 2006, \aap, 460, 339

\bibitem[{{Joshi} {et~al.}(2014){Joshi}, {Srivastava}, {Banerjee},
  {Venkataraman}, \& {Ashok}}]{joshi14}
{Joshi}, V., {Srivastava}, M., {Banerjee}, D.~P.~K., {Venkataraman}, V., \&
  {Ashok}, N.~M. 2014, ATEL, 6772, 1

\bibitem[{Kasliwal {et~al.}(2014)Kasliwal, Tinyanont, Jencson, Cao, Perley,
  O'Sullivan, Prince, Masci, Helou, Armus, Surace, Cody, Dyk, Bond, Bally,
  Levesque, Williams, Whitelock, Mohamed, Gehrz, Shenoy, Carlon, Corgan,
  Dykhoff, Smith, Cantiello, Langer, Ofek, Parthasarathy, Phillips, Hsiao,
  Morrell, Gonzalez, \& Contreras}]{Kasliwal:2014p24264}
Kasliwal, M.~M., Tinyanont, S., Jencson, J., {et~al.} 2014, ATEL, 6644, 1

\bibitem[{Khokhlov(1991)}]{khokhlov91}
Khokhlov, A.~M. 1991, A\&A, 245, 114

\bibitem[{Maoz {et~al.}(2014)Maoz, Mannucci, \& Nelemans}]{maoz13}
Maoz, D., Mannucci, F., \& Nelemans, G. 2014, ARA\&A, 52, 107

\bibitem[{Markwardt(2009)}]{markwardt09}
Markwardt, C.~B. 2009, in Astronomical Data Analysis Software and Systems
  XVIII, ed. D.~A. Bohlender, D.~Durand, \& P.~Dowler (San Francisco: ASP), 251

\bibitem[{Marshall {et~al.}(2004)Marshall, van Loon, Matsuura, Wood, Zijlstra,
  \& Whitelock}]{marshall04}
Marshall, J.~R., van Loon, J.~T., Matsuura, M., {et~al.} 2004, MNRAS, 355, 1348

\bibitem[{{McCully} {et~al.}(2014){McCully}, {Jha}, {Foley}, {Bildsten},
  {Fong}, {Kirshner}, {Marion}, {Riess}, \& {Stritzinger}}]{mccully14}
{McCully}, C., {Jha}, S.~W., {Foley}, R.~J., {et~al.} 2014, \nat, 512, 54

\bibitem[{Nakano {et~al.}(2014)Nakano, Itagaki, Guido, Nicolini, Howes, Kiyota,
  Masi, Catalano, Vagnozzi, \& Munari}]{nakano14}
Nakano, S., Itagaki, K., Guido, E., {et~al.} 2014, CBET, 4011, 1

\bibitem[{Nozawa {et~al.}(2011)Nozawa, Maeda, Kozasa, Tanaka, Nomoto, \&
  Umeda}]{nozawa11}
Nozawa, T., Maeda, K., Kozasa, T., {et~al.} 2011, ApJ, 736, 45

\bibitem[{Ochner {et~al.}(2014)Ochner, Tomasella, Benetti, Cappellaro,
  Elias-Rosa, Pastorello, \& Turatto}]{ochner14}
Ochner, P., Tomasella, L., Benetti, S., {et~al.} 2014, CBET, 4011, 2

\bibitem[{Penney \& Hoeflich(2014)}]{Penney:2014p24317}
Penney, R., \& Hoeflich, P. 2014, ApJ, 795, 84

\bibitem[{Phillips(1993)}]{phillips93}
Phillips, M.~M. 1993, ApJ, 413, L105

\bibitem[{Phillips {et~al.}(2007)Phillips, Li, Frieman, Blinnikov, DePoy,
  Prieto, Milne, Contreras, Folatelli, Morrell, Hamuy, Suntzeff, Roth,
  Gonz{\'a}lez, Krzeminski, Filippenko, Freedman, Chornock, Jha, Madore,
  Persson, Burns, Wyatt, Murphy, Foley, Ganeshalingam, Serduke, Krisciunas,
  Bassett, Becker, Dilday, Eastman, Garnavich, Holtzman, Kessler, Lampeitl,
  Marriner, Frank, Marshall, Miknaitis, Sako, Schneider, van~der Heyden, \&
  Yasuda}]{philips07}
Phillips, M.~M., Li, W., Frieman, J.~A., {et~al.} 2007, PASP, 119, 360

\bibitem[{{Rodr{\'{\i}}guez} {et~al.}(2014){Rodr{\'{\i}}guez}, {Clocchiatti},
  \& {Hamuy}}]{2014AJ....148..107R}
{Rodr{\'{\i}}guez}, {\'O}., {Clocchiatti}, A., \& {Hamuy}, M. 2014, \aj, 148,
  107

\bibitem[{{Roming} {et~al.}(2005){Roming}, {Kennedy}, {Mason}, {Nousek}, {Ahr},
  {Bingham}, {Broos}, {Carter}, {Hancock}, {Huckle}, {Hunsberger}, {Kawakami},
  {Killough}, {Koch}, {McLelland}, {Smith}, {Smith}, {Soto}, {Boyd},
  {Breeveld}, {Holland}, {Ivanushkina}, {Pryzby}, {Still}, \&
  {Stock}}]{2005SSRv..120...95R}
{Roming}, P.~W.~A., {Kennedy}, T.~E., {Mason}, K.~O., {et~al.} 2005, \ssr, 120,
  95

\bibitem[{{Roy} {et~al.}(2011){Roy}, {Kumar}, {Benetti}, {Pastorello}, {Yuan},
  {Brown}, {Immler}, {Fatkhullin}, {Moskvitin}, {Maund}, {Akerlof}, {Wheeler},
  {Sokolov}, {Quimby}, {Bufano}, {Kumar}, {Misra}, {Pandey}, {Elias-Rosa},
  {Roming}, \& {Sagar}}]{2011ApJ...736...76R}
{Roy}, R., {Kumar}, B., {Benetti}, S., {et~al.} 2011, \apj, 736, 76

\bibitem[{{Sahu} {et~al.}(2008){Sahu}, {Tanaka}, {Anupama}, {Kawabata},
  {Maeda}, {Tominaga}, {Nomoto}, {Mazzali}, \& {Prabhu}}]{2008ApJ...680..580S}
{Sahu}, D.~K., {Tanaka}, M., {Anupama}, G.~C., {et~al.} 2008, \apj, 680, 580

\bibitem[{{Schoeniger} \& {Sofue}(1997)}]{1997A&A...323...14S}
{Schoeniger}, F., \& {Sofue}, Y. 1997, \aap, 323, 14

\bibitem[{Shen {et~al.}(2013)Shen, Guillochon, \& Foley}]{shen13}
Shen, K.~J., Guillochon, J., \& Foley, R.~J. 2013, ApJL, 770, L35

\bibitem[{Silverman {et~al.}(2013)Silverman, Nugent, Gal-Yam, Sullivan, Howell,
  Filippenko, Arcavi, Ben-Ami, Bloom, Cenko, Cao, Chornock, Clubb, Coil, Foley,
  Graham, Griffith, Horesh, Kasliwal, Kulkarni, Leonard, Li, Matheson, Miller,
  Modjaz, Ofek, Pan, Perley, Poznanski, Quimby, Steele, Sternberg, Xu, \&
  Yaron}]{silverman13b}
Silverman, J.~M., Nugent, P.~E., Gal-Yam, A., {et~al.} 2013, ApJS, 207, 3

\bibitem[{Soker {et~al.}(2013)Soker, Kashi, Garc{\'\i}a-Berro, Torres, \&
  Camacho}]{soker13}
Soker, N., Kashi, A., Garc{\'\i}a-Berro, E., Torres, S., \& Camacho, J. 2013,
  MNRAS, 431, 1541

\bibitem[{Stritzinger {et~al.}(2015)Stritzinger, Valenti, Hoeflich, Baron,
  Phillips, Taddia, Foley, Hsiao, Jha, McCully, Pandya, Simon, Benetti, Brown,
  Burns, Campillay, Contreras, F{\"o}rster, Holmbo, Marion, Morrell, \&
  Pignata}]{stritzinger15}
Stritzinger, M.~D., Valenti, S., Hoeflich, P., {et~al.} 2015, A\&A, 573, A2

\bibitem[{Szalai \& Vink{\'o}(2013)}]{szalai13}
Szalai, T., \& Vink{\'o}, J. 2013, A\&A, 549, 79

\bibitem[{Valenti {et~al.}(2009)Valenti, Pastorello, Cappellaro, Benetti,
  Mazzali, Manteca, Taubenberger, Elias-Rosa, Ferrando, Harutyunyan, Hentunen,
  Nissinen, Pian, Turatto, Zampieri, \& Smartt}]{valenti09}
Valenti, S., Pastorello, A., Cappellaro, E., {et~al.} 2009, Nature, 459, 674

\bibitem[{{Valenti} {et~al.}(2011){Valenti}, {Fraser}, {Benetti}, {Pignata},
  {Sollerman}, {Inserra}, {Cappellaro}, {Pastorello}, {Smartt}, {Ergon},
  {Botticella}, {Brimacombe}, {Bufano}, {Crockett}, {Eder}, {Fugazza},
  {Haislip}, {Hamuy}, {Harutyunyan}, {Ivarsen}, {Kankare}, {Kotak}, {Lacluyze},
  {Magill}, {Mattila}, {Maza}, {Mazzali}, {Reichart}, {Taubenberger},
  {Turatto}, \& {Zampieri}}]{2011MNRAS.416.3138V}
{Valenti}, S., {Fraser}, M., {Benetti}, S., {et~al.} 2011, \mnras, 416, 3138

\bibitem[{Werner {et~al.}(2004)Werner, Roellig, Low, Rieke, Rieke, Hoffmann,
  Young, Houck, Brandl, Fazio, Hora, Gehrz, Helou, Soifer, Stauffer, Keene,
  Eisenhardt, Gallagher, Gautier, Irace, Lawrence, Simmons, Cleve, Jura,
  Wright, \& Cruikshank}]{Werner:2004p24304}
Werner, M.~W., Roellig, T.~L., Low, F.~J., {et~al.} 2004, ApJS, 154, 1

\bibitem[{White {et~al.}(2015)White, Kasliwal, Nugent, Gal-Yam, Howell,
  Sullivan, Goobar, Piro, Bloom, Kulkarni, Laher, Masci, Ofek, Surace, Ben-Ami,
  Cao, Cenko, Hook, J{\"o}nsson, Matheson, Sternberg, Quimby, \&
  Yaron}]{white15}
White, C.~J., Kasliwal, M.~M., Nugent, P.~E., {et~al.} 2015, ApJ, 799, 52

\bibitem[{Williams \& Fox(2015)}]{williams15}
Williams, B.~J., \& Fox, O.~D. 2015, ApJL, 808, L22

\bibitem[{{Yamanaka} {et~al.}(2015){Yamanaka}, {Maeda}, {Kawabata}, {Tanaka},
  {Tominaga}, {Akitaya}, {Nagayama}, {Kuroda}, {Takahashi}, {Saito},
  {Yanagisawa}, {Fukui}, {Miyanoshita}, {Watanabe}, {Arai}, {Isogai},
  {Hattori}, {Hanayama}, {Itoh}, {Ui}, {Takaki}, {Ueno}, {Yoshida}, {Ali},
  {Essam}, {Ozaki}, {Nakao}, {Hamamoto}, {Nogami}, {Morokuma}, {Oasa},
  {Izumiura}, \& {Sekiguchi}}]{2015ApJ...806..191Y}
{Yamanaka}, M., {Maeda}, K., {Kawabata}, K.~S., {et~al.} 2015, \apj, 806, 191

\end{thebibliography}

\end{document}